\documentclass[conference]{IEEEtran}
\IEEEoverridecommandlockouts
\usepackage{cite}
\usepackage{amsmath,amssymb,amsfonts}
\usepackage{algorithmic}
\usepackage{graphicx}
\usepackage{textcomp}
\usepackage{xcolor}
\usepackage{url}
\usepackage{amsmath}
\usepackage{mathtools}
\usepackage{threeparttable}
\usepackage{booktabs}
\usepackage{float}
\usepackage{cite,balance}
\usepackage{subfigure}
\def\BibTeX{{\rm B\kern-.05em{\sc i\kern-.025em b}\kern-.08em
    T\kern-.1667em\lower.7ex\hbox{E}\kern-.125emX}}

\begin{document}

\title{Transformer-based Arabic Dialect Identification}
\author{\IEEEauthorblockN{Wanqiu Lin, Maulik Madhavi, Rohan Kumar Das and Haizhou Li}
\IEEEauthorblockA{\textit{Department of Electrical and Computer Engineering,} \\
\textit{National University of Singapore, Singapore} \\
e0506986@u.nus.edu,\{maulik.madhavi, rohankd, haizhou.li\}@nus.edu.sg}
}

\maketitle

\begin{abstract}
This paper presents a dialect identification (DID) system based on the transformer neural network architecture. The conventional convolutional neural network (CNN)-based systems use the shorter receptive fields. We believe that long range information is equally important for language and DID, and self-attention mechanism in transformer captures the long range dependencies. In addition, to reduce the computational complexity, self-attention with downsampling is used to process the acoustic features. This process extracts sparse, yet informative features. Our experimental results show that transformer outperforms CNN-based networks on the Arabic dialect identification (ADI) dataset. We also report that the score-level fusion of CNN and transformer-based systems obtains an overall accuracy of 86.29\% on the ADI17 database. 
\end{abstract}

\vspace{2mm}

\section{Introduction}

Dialect identification (DID) is the task to identify different dialects within the same language family and can be considered as a special topic of language identification (LID). However, DID is generally more challenging than common LID tasks, since the similar dialects often share the close feature space, such as the acoustic, linguistic, and speaker characteristics. The valuable outcomes from the DID are useful to train the language and dialect-specific acoustic model for automatic speech recognition (ASR). Thus, DID is an essential technology in voice-interactive smart devices, such as Google Home, Alexa, Siri, etc.

The  Multi-Genre Broadcast (MGB) challenge\footnote{http://www.mgb-challenge.org} is a series of evaluations of several speech technologies, which include speaker diarization,  speech recognition, DID and lightly supervised alignment using multimedia database\cite{ali2019mgb}. The regional Arabic dialects capture the common base of character sets and phonetic inventory, whereas the dialects are mutually unintelligible. This makes DID a challenging research problem. Arabic dialect identification-17  (ADI17) \footnote{https://arabicspeech.org/mgb5} is one of the two tracks in the latest MGB-5 challenge. The task aims to identify one of 17 Arabic dialects from the speech audios collected from YouTube. In this paper, we use the ADI17 dataset for DID task\footnote{http://groups.csail.mit.edu/sls/downloads/adi17}.


In the literature, it is observed that longer temporal information contributes to phonotactic information, which is very important for language and dialect recognition~\cite{torresapproaches,tong2006integrating,IFCC_icassp}. In particular, several research studies use phone recognition followed by the language model technique to use phonotactic information for language identification task \cite{tong2006integrating,LiM05,SrivastavaVVS17}. Similarly, long range information based features are found to be useful for detection tasks~\cite{rkd_is2019,Das2019,rkd_ASRU2019}. Along a similar direction, lexical features are used together with audio features to formulate a vector-space model or n-gram for the Arabic DID task \cite{shon2018convolutional,KhuranaNAHBG17}. However, in a multi-dialect scenario, automatic speech recognition is far from perfect \cite{ali2019mgb}. Furthermore, the absence of unified orthographic rules or Arabic language makes it difficult to train and evaluate \cite{ali2019mgb,AliM0R15}.  

\subsection{Related Work in DID}
Recent studies show that end-to-end DID consist convolutional neural network (CNN) and residual network (ResNet) for DID improve performance over the traditional approaches  \cite{shon2018convolutional,miao2019new,miao2019lstm,cai2019utterance}. These studies indicate that there is scope for further detailed analyses of end-to-end frameworks on DID research. In this context, we believe that the transformer model can be useful as it has the ability to learn the temporal sequence information with no sequential operation in execution \cite{vaswani2017attention}. In addition, the longer temporal information captured using the self-attention mechanism in the transformer may help the DID task. With this motivation, we propose an end-to-end DID system using a transformer model in this paper. To reduce the computation requirement in self-attention, we use a downsampling approach to reduce the length of sequence and thereby the number of operations.


The rest of the paper is organized as follows: Section \ref{section2} describes the structure of two end-to-end DID models including the acoustic features we examined as well as the processing method to support consideration of Transformer in this work. The details of the experimental setup are discussed in Section \ref{section3}. Section \ref{section4} describes the experimental results with analysis. Finally,  Section \ref{section5} concludes the work with future research directions.

\section{Transformer-based DID system}
\label{section2}
Next we formulate a transformer based DID system. We first describe the neural network (NN) architecture and computational challenges associated with the self-attention layer. The architecture of the transformer is widely popular due to the self-attention operation that does not use recurrent, and convolution operations \cite{vaswani2017attention}. The self-attention mechanism can relate and learn the dependencies along a longer speech sequence. The transformer was initially proposed on the machine translation, and further applied to speech recognition. Several studies used the transformer model to perform acoustic modeling. In this paper, we follow the speech-transformer model described in~\cite{li2019speechtransformer}. 
In the DID task, we only use the encoder components from the speech transformer model as a feature extractor. It is to be noted that our objective is to identify the dialect, which does not fall under sequence to sequence problems such as speech recognition.  

\begin{table}[t!]
\caption{Computational cost for self-attention and convolutional layers ($n$= sequence length, $d$ = the dimension of input representation, $k$ = kernel size used in convolutions)\cite{vaswani2017attention}}
\begin{center}
\begin{tabular}{|c|c|c|}
\hline
{\bf Layer Type}                                                                   & {\bf Complexity/Layer}    & {\bf Maximum Path Length} \\ \hline\hline
Self-Attention                                                               & $O(n^{2}\cdot d) $                   & $O(1)$              \\ \hline
\begin{tabular}[c]{@{}c@{}}Convolutional\\ (contiguous kernels)\end{tabular} & $O(k\cdot n\cdot d^{2})$  & $O(n/k)$            \\ \hline
\end{tabular}
\label{table:why attention}
\end{center}
\vspace{-2mm}
\end{table}

\subsection{Computational Issues in Transformer}\label{section14}
Self-attention mechanism represents the core computing unit in transformer model, that processes every input speech frame. For speech, the number of frames is much larger than the input dimension, thus it requires more computation than the conventional CNN.  Assuming that transformer performs mapping from one sequence $\{x_1,...,x_n\} $ to another sequences having same length $\{z_1,...,z_n\} $, such that $x_i,z_i\in R_d$. The comparison of complexity in self-attention layers and convolution layers is shown in Table \ref{table:why attention}.  Both the convolutional layer and the self-attention layer have a constant number of sequentially executed operations \cite{vaswani2017attention}. Therefore, both of them can perform parallel computation that recurrent networks cannot. However, it is found that the computational complexity is related to quadratic order of the length, i.e., $\mathcal{O}(n^2 d)$ where $n$ and $d$ are the length of the input sequence and dimension of input, respectively. This may not be important for text sequence in natural language processing (NLP) tasks as text sequence length is lesser than the length of speech sequence \cite{vaswani2017attention}. Moreover, the adjacent frames in speech are more correlated than the adjacent words in text sequence. 

There have been studies to address this sequence length problem in transformer architecture. Some studies used restricted context to compute the self-attention rather than using entire audio length \cite{PoveyHGLK18,transformer_transducer}. Downsampling and frame reduction are such examples \cite{PhamNN0W19,SperberNNSW18,dong2018speech,li2019speechtransformer}. Similarly, the use of subsampling and pooling is presented in \cite{SalazarKH19} for ASR. Others describe the use of convolutional layers to perform downsampling \cite{AlexBie,TsunooKKW19}. For recurrent architecture-based ASR, some studies also follow the similar approaches known as Low Frame Rate (LFR) \cite{sak2015fast,sak2015acoustic,pundak2016lower}. In this context, we follow the method in \cite{li2019speechtransformer} that performs stacking and subsampling to reduce the number of frames to overcome the computational issue.

We believe that the introduction of downsampling technique into DID has two advantages. Firstly, since the speech feature sequence is longer than the word sequence. When we introduced the transformer model from machine translation to DID,
reducing the frames rate can significantly improve the calculation efficiency. Secondly, there is no clear boundary between the speech frames, which may be difficult for the encoder layers of the speech transformer to calculate the similarity of adjacent frames. After the acoustic features are stacked and down-sampled, acoustic features will produce more sparse but useful information that is beneficial to DID.

\subsection{Neural Network Architecture}\label{subsection13}

\begin{figure}
    \centering
    \includegraphics[trim=20 0 0 0,clip=true,scale=0.4]{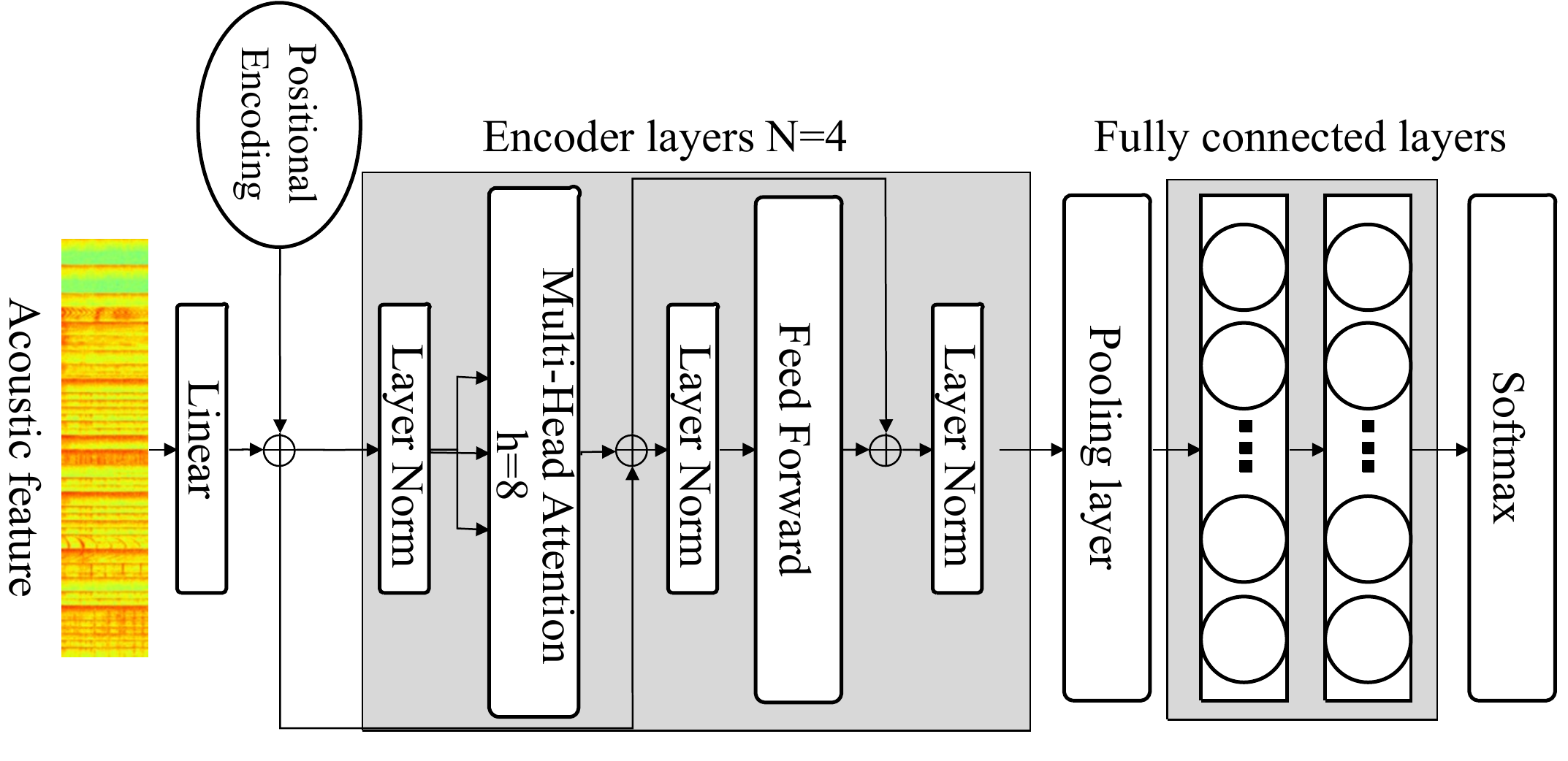}
    \caption{Architecture of transformer-based DID system.}
    \label{fig:TransformerFC}
\end{figure}

The structure of the transformer-based DID system is shown in Fig. \ref{fig:TransformerFC}.
As discussed earlier, there is a need to modify the sequence length to reduce the complexity. First, we perform downsampling to reduce the number of input lengths as suggested in \cite{li2019speechtransformer}. Then, to preserve the positional information in the sequence, we add positional encoding representation to the input encoding. There are many options for position coding \cite{positional_encoding,gehring2017convolutional}. Here, we choose sinusoidal functions to obtain the positional encoding \cite{vaswani2017attention}:
\begin{eqnarray}
PE_{(pos,2i)} =\sin(pos/10000^{2i/d_{model}}) \\
PE_{(pos,2i+1)} =\cos(pos/10000^{2i/d_{model}}) 
\end{eqnarray}
where $pos$ and $i$ represent the position and dimension, respectively. The encoder is composed of a stack of $N=4$ identical layers, where each layer contains two sublayers. The first sublayer does multi-head self-attention (MHA) operation, while the second sublayer is a position-wise fully connected feed-forward network. The residual connections are used around each sublayers, and it is followed by a layer normalization operation. For MHA, the number of heads $h=8$ are used. The MHA allows the model to focus on information from different acoustic representation, which is used in all sublayers and embedding layers in the model produce an output having dimension, $d_{model} = 512$. 
 As showed in Fig. \ref{fig:TransformerFC}, the output of encoder layers are taken by a global average pooling layer to calculate the mean and and standard deviation, and obtain a fixed dimension representation.  After global average pooling, the fixed-length output is then given to two additional fully connected layers with 512 and 64 nodes, respectively and followed by classification layer with 17 nodes corresponds to the number of dialects. 

\section{Experimental Setup}
\label{section3}
\subsection{Dataset}\label{subsection21}

The ADI17 dataset is provided by the MGB5 organization, which contains 17 dialects of Arab countries. These dialects are collected from YouTube \cite{ShonASMG20,ali2019mgb}. The training set contains about 3,000 hours of data, and the test set and dev set are about 280 hours of data. As these data are collected in the channel of a specific country, it may contain some label errors, which is conducive to unsupervised learning. After manual calibration and verification, 57 hours of data is selected as the test set and dev set for performance evaluation. At the same time, these data are divided into three testing sub-categories according to the duration of the segments: short duration ($<$ 5s), medium duration (5-20s), and long duration ($>$20s).
More detailed information about the dataset is available in \cite{ali2019mgb}. It is also observed that the training set of ADI is very unbalanced in terms of the number of train spoken utterances per dialect. The Iraq (IRA) dialect has a large number (291,123) of utterances. On the other hand, the Jordanian (JOR) dialect has a very low number (5514) of utterances. 

\subsection{Experimental Parameters}
\label{subsection22}
In this paper, we use the Kaldi\footnote{https://kaldi-asr.org/} toolkit \cite{povey2011kaldi} to extract 80-dimensional Fbank features from Arabic dialect audio. The features are extracted with 25 ms frame-length and 10 ms frame-shift. Then cepstral mean and variance normalization is performed to the original features.
We applied the down-sampling approach as discussed earlier to process the speech sequence. For implementation, we set the down-sampling factor is $n = 3$, and the stacking factor is $m = 4$ as that in \cite{li2019speechtransformer}. This will reduce the number of frames by 3 and increased the input dimension by 4, i.e., 320 from 80 dimensional Fbank features. 
 
\begin{figure}[t!]
    \begin{center}
    \includegraphics[trim=20 0 0 0,clip=true,scale=0.4]{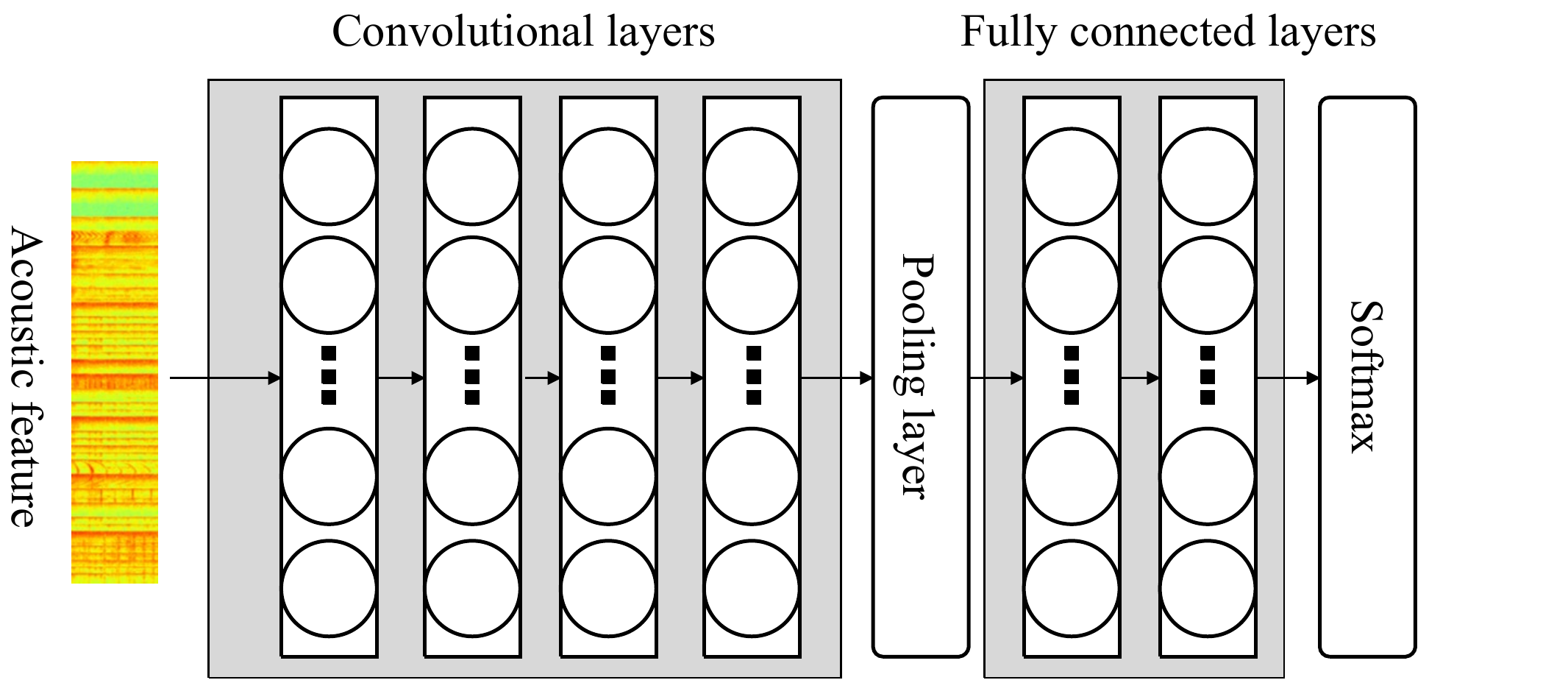}
    \caption{Architecture of CNN-based DID system}
    \label{fig:CNNFC}        
    \end{center}
\end{figure}

The transformer-based model contains 4 encoder layers, which are configured as $h=8$ attention heads, $d\_{model} = 512$ model dimension and $d\_{inner} = 2048$ inner-layer dimension, followed by a pooling layer and two fully connected layers, with the hidden nodes 512 and 64, respectively. We use stochastic gradient descent (SGD) optimizer with a momentum of 0.8. The initial value of the learning rate is set to 0.001, and we reduce the learning rate whenever the validation accuracy plateaus. In the training stage, we set the mini-batch size at 10. The CNN-based network has the same configuration as \cite{ali2019mgb}. The structure of the CNN-based DID system is shown in Fig. \ref{fig:CNNFC}. 
The output from softmax can be used directly as a score for each Arabic dialect. The performance is measured in terms of \% accuracy. Next, we present the experimental results using a transformer-based DID system. The source code is available at \footnote{\url{https://github.com/LIN-WANQIU/ADI17}}.

\begin{table*}[htbp]
\centering
\caption{Performance results for transformer-based DID system with feature reduction by downsampling}
\begin{tabular}{@{}ccccccccc@{}}
\toprule
    & \multicolumn{4}{c}{\bf Dev}               & \multicolumn{4}{c}{\bf Test}              \\ \midrule
{\bf Downsampling} &{\bf  \textless{}5sec} & {\bf 5sec$\sim$20sec} & {\bf \textgreater{}20sec} & {\bf Overall} & {\bf  \textless{}5sec} & {\bf 5sec$\sim$20sec} & {\bf \textgreater{}20sec} & {\bf Overall}  \\ \midrule
Yes & 75.97\% & 86.39\% & 92.17\% & 83.17\% & 76.21\% & 86.01\% & 90.58\% & 82.54\% \\ \midrule
No  & 66.87\% & 78.09\% & 83.06\% & 74.51\% & 69.43\% & 79.27\% & 83.13\% & 75.97\% \\ \bottomrule
\end{tabular}
\label{table:resultmLFR}
\end{table*}

\section{Experimental Results}
\label{section4}
It can be seen from Table \ref{table:resultmLFR} that downsampling contributes to about 7\% accuracy improvement. This may be because the adjacent frames in a speech before downsampling are highly correlated and produce redundant information, which does not provide essential information needed in the DID task. Using this downsampling approach, the self-attention mechanism of the encoder can be more effective by reducing the length of the speech sequence and thus speed up the training task. The speech sequence processed by the downsampling approach can generate more useful information, which is beneficial to our dialect recognition. When using downsampling, the real-time factor (RTF) of the system can be accelerated by about 1.5 times as the number of frames reduced by 3.
\begin{table*}[htbp]
\centering
\caption{Performance results for CNN and transformer-based DID system}
\begin{tabular}{@{}ccccccccc@{}}
\toprule
                   & \multicolumn{4}{c}{\bf Dev}               & \multicolumn{4}{c}{\bf Test}              \\ \midrule
{\bf Systems} & {\bf  \textless{}5sec} & {\bf 5sec$\sim$20sec} & {\bf \textgreater{}20sec} & {\bf Overall}  & {\bf  \textless{}5sec} & {\bf 5sec$\sim$20sec} & {\bf \textgreater{}20sec} & {\bf Overall}  \\ \midrule
CNN         & 78.17\% & 82.77\% & 89.72\% & 78.17\% & 68.36\% & 83.34\% & 87.10\% & 77.77\% \\ \midrule
Transformer  & 75.97\% & 86.39\% & 92.17\% & 83.17\% & 76.21\% & 86.01\% & 90.58\% & 82.54\%\\\midrule
Fusion & 78.19\%         & 88.41\%         & 94.04\% & 85.25\%  & 80.95\%         & 89.23\%         & 93.06\% & 86.29\%\\\bottomrule
\end{tabular}
\label{table:resultCNNTransformer}
\end{table*}


\begin{figure*}[!tbp]
\begin{center}
 \includegraphics[trim=60 200 60 170,clip=true,scale=0.65]{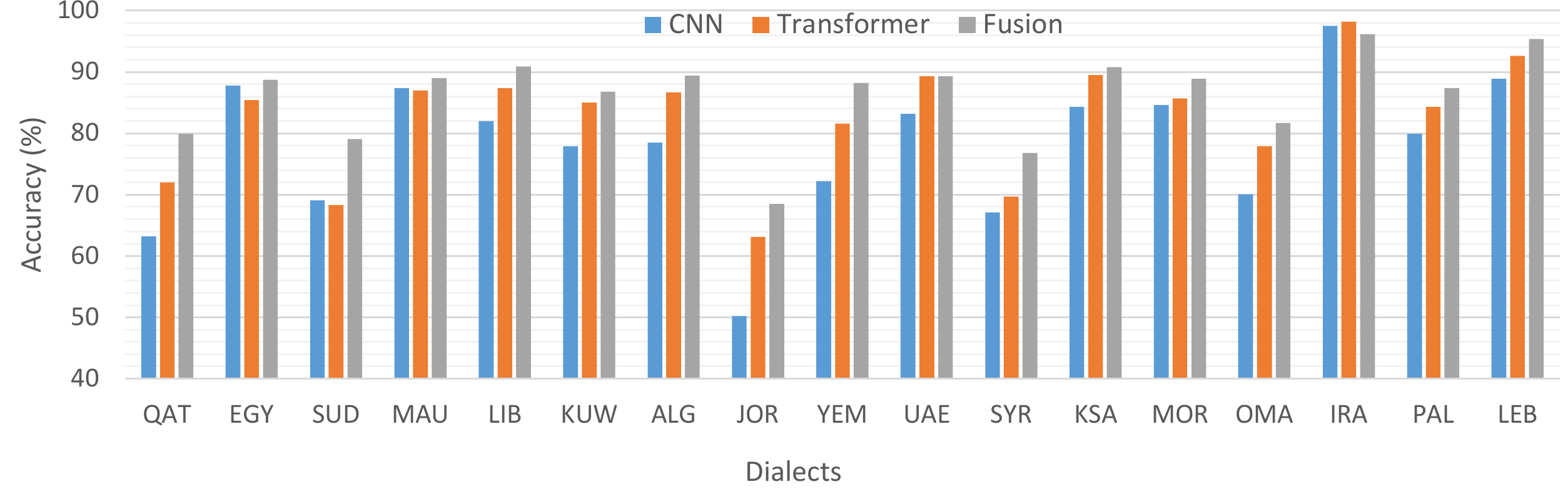} 
\end{center}
\vspace{-4mm}
\caption{Effect of score-level fusion of CNN and transformer-based DID system}
\label{figure:Transformer_CNN}
\vspace{-2mm}
\end{figure*}

Table \ref{table:resultCNNTransformer} shows the results of the CNN-based and transformer-based DID systems. 
We can observe that our transformer-based DID system is slightly better than the baseline in terms of overall accuracy. To compare with similar acoustic features, we also experimented with the CNN based system trained on Fbank features. We observe relatively better performance by transformer than CNN. 
Especially, the transformer-based DID system is better in the cases for medium and long duration. 
For the shorter duration ($<5$s) of test speech utterance does not provide more discriminatory features for dialect identification. This is due to the downsampling used while input feature processing for transformer-based DID system. In addition, it is commonly observed across different systems that as duration of test speech increases, the performance in accuracy improves. The results show that our model is slightly better than the CNN and has an accuracy of about 4-5\% higher than the CNN model. Although the transformer model uses the attention mechanism to capture global information and can improve the learning ability of long-term dependencies, the consideration of the relation between the position of the input sequence is relatively simple, so the performance on the long sequence is not as good as expected. It indicates that studying the nature of the position sequence of sentences in the transformer may be a very important direction, which can further improve the performance of the transformer model.


Next, we conduct an experiment on the score-level fusion of two systems. We averaged the scores obtained from the classification layers of CNN and transformer systems. The results of score-level fusion are shown in Table \ref{table:resultCNNTransformer}. It is observed that the overall accuracy of the fusion system is 86.29\% that outperforms the baseline system used in \cite{ali2019mgb}. From Table \ref{table:resultCNNTransformer}, we perform further investigations into the error patterns for utterances of different duration. It is observed that the longer duration of data leads to improved performance. 

The overall accuracy after score-level fusion is increased by 3-4\% as compared to the transformer-based system. To understand the confusion across different dialects,  we further analyzed confusion across different miss classifications. Fig.  \ref{figure:Transformer_CNN} shows the DID performance for each dialect. The performance is shown for all three systems, namely, CNN, transformer and the score-level fusion system. It is observed that score-level fusion gives better performance in terms of accuracy for the majority of all 17 dialects. We can observe that the classification accuracy for most of the dialect is above 85\% except for JOR, QAT and QMA. Among all the 17 dialects, the IRA obtains 96.2\% accuracy which is the best performance, followed by LEB dialect with an accuracy of 95.4\%.  It is easy to notice that the dialect JOR is the most confusing dialect, with an accuracy of only 68.5\%, and is most often confused with PAL (6.1\%) and LEB (5.1\%). 
The second confusing dialect is SYR, which has an accuracy of 76.8\% and is most often confused with EGY (5.7\%). As discussed earlier in section \ref{subsection21} that ADI data has unbalanced nature in terms of utterances available.  It is also observed that performance of dialect identification is affected by the number of training utterances used. It was observed that the IRA dialect has relatively better performance while the JOR has poor performance. The IRA dialect has a larger number of utterances, while the JOR dialect has the least number of utterances.

\begin{table}[ht]
\caption{Comparisons of different methods for ADI}
\begin{center}
\begin{tabular}{|c|c|c|}
\hline
{\bf Systems}                                                                   & {\bf Dev}    & {\bf Eval} \\ \hline \hline
i-vector \cite{ShonASMG20}  &    59.7\%     &    60.3\%          \\ \hline
x-vector \cite{ShonASMG20}  &    71.0\%    &       72.1\%   \\ \hline
E2E (x-vector) \cite{ShonASMG20}  &  76.6\%  &     77.8\%         \\ \hline
E2E (Softmax) \cite{ShonASMG20}  &   83.0\%     &  82.0\%            \\ \hline
E2E (Tuplemax)  \cite{ShonASMG20} &  78.6\%  & 78.6\%             \\ \hline
E2E (AM-Softmax)  \cite{ShonASMG20} & 62.5\%    &  63.7\%             \\ \hline
Transformer & 83.17\% & 82.54\% \\ \hline
CNN-Transformer (fusion) & \textbf{85.25}\% & \textbf{86.29}\% \\ \hline 
\end{tabular}
\label{table:different_methods}
\end{center}
\vspace{-2mm}
\end{table} 
Table \ref{table:different_methods} shows the experimental results with different state-of-the art DID systems. The systems such as i-vector and x-vector are inspired by speaker verification and use logistic regression for scoring. These are not trained as an end-to-end (E2E) manner and are not as good as E2E systems. transformer-based DID system presented in this paper is comparable with E2E (softmax). We believe that the longer temporal information captured using the self-attention mechanism in transformer helps in improving the performance of DID system.

\section{Conclusions}
\label{section5}
In this paper, we proposed a transformer-based DID system. We observed that long-term information captured using the transformer model can help in DID performance. To lower the computational cost by reducing the number of frames for self-attention layer, we used the downsampling based approach. This approach also eliminated the redundant adjacent Fbank features. We observed relatively better performance transformer-based DID system than CNN-based DID system for medium and long duration of test speech. This is expected as the transformer has the capability to learn the long time dependencies, where as CNN-based system has shorter convolution filters. Further, we perform the score-level fusion of CNN and transformer-based DID systems to analyse the DID performance for different dialects. The future work will focus on using additional resources such as ASR, lexical information and phonotactic modeling approaches together with audio data for the DID task. 


\balance
\bibliographystyle{IEEEtran}  
\bibliography{reference.bib}  

\begin{thebibliography}{10}
\providecommand{\url}[1]{#1}
\csname url@samestyle\endcsname
\providecommand{\newblock}{\relax}
\providecommand{\bibinfo}[2]{#2}
\providecommand{\BIBentrySTDinterwordspacing}{\spaceskip=0pt\relax}
\providecommand{\BIBentryALTinterwordstretchfactor}{4}
\providecommand{\BIBentryALTinterwordspacing}{\spaceskip=\fontdimen2\font plus
\BIBentryALTinterwordstretchfactor\fontdimen3\font minus
  \fontdimen4\font\relax}
\providecommand{\BIBforeignlanguage}[2]{{%
\expandafter\ifx\csname l@#1\endcsname\relax
\typeout{** WARNING: IEEEtran.bst: No hyphenation pattern has been}%
\typeout{** loaded for the language `#1'. Using the pattern for}%
\typeout{** the default language instead.}%
\else
\language=\csname l@#1\endcsname
\fi
#2}}
\providecommand{\BIBdecl}{\relax}
\BIBdecl

\bibitem{ali2019mgb}
A.~Ali, S.~Shon, Y.~Samih, H.~Mubarak, A.~Abdelali, J.~Glass, S.~Renals, and
  K.~Choukri, ``The {MGB-5} challenge: Recognition and dialect identification
  of dialectal arabic speech,'' in \emph{Automatic Speech Recognition and
  Understanding Workshop (ASRU)}, 2019, pp. 1026--1033.

\bibitem{torresapproaches}
P.~A. Torres{-}Carrasquillo, E.~Singer, M.~A. Kohler, R.~J. Greene, D.~A.
  Reynolds, and J.~R.~D. Jr., ``Approaches to language identification using
  {G}aussian mixture models and shifted delta cepstral features,'' in \emph{7th
  International Conference on Spoken Language Processing, {ICSLP2002} -
  {INTERSPEECH}}, J.~H.~L. Hansen and B.~L. Pellom, Eds.\hskip 1em plus 0.5em
  minus 0.4em\relax {ISCA}, 2002.

\bibitem{tong2006integrating}
R.~Tong, B.~Ma, D.~Zhu, H.~Li, and E.~Chng, ``Integrating acoustic, prosodic
  and phonotactic features for spoken language identification,'' in
  \emph{International Conference on Acoustics, Speech and Signal Processing
  {(ICASSP)}}, 2006, pp. 205--208.

\bibitem{IFCC_icassp}
K.~Vijayan, H.~Li, H.~Sun, and K.~A. Lee, ``On the importance of analytic phase
  of speech signals in spoken language recognition,'' in \emph{International
  Conference on Acoustics, Speech and Signal Processing {(ICASSP)}}, 2018, pp.
  5194--5198.

\bibitem{LiM05}
H.~Li and B.~Ma, ``A phonotactic language model for spoken language
  identification,'' in \emph{{ACL} 2005, 43rd Annual Meeting of the Association
  for Computational Linguistics, Proceedings of the Conference, 25-30 June
  2005, University of Michigan, {USA}}, K.~Knight, H.~T. Ng, and K.~Oflazer,
  Eds.\hskip 1em plus 0.5em minus 0.4em\relax The Association for Computer
  Linguistics, 2005, pp. 515--522.

\bibitem{SrivastavaVVS17}
B.~M.~L. Srivastava, H.~K. Vydana, A.~K. Vuppala, and M.~Shrivastava,
  ``Significance of neural phonotactic models for large-scale spoken language
  identification,'' in \emph{International Joint Conference on Neural Networks,
  {IJCNN}}, 2017, pp. 2144--2151.

\bibitem{rkd_is2019}
R.~K. Das, J.~Yang, and H.~Li, ``Long range acoustic features for spoofed
  speech detection,'' in \emph{Interspeech}, 2019, pp. 1058--1062.

\bibitem{Das2019}
R.~K. Das and H.~Li, ``Instantaneous phase and long-term acoustic cues for orca
  activity detection,'' in \emph{Interspeech}, 2019, pp. 2418--2422.

\bibitem{rkd_ASRU2019}
R.~K. Das, J.~Yang, and H.~Li, ``Long range acoustic and deep features
  perspective on {ASV}spoof 2019,'' in \emph{Automatic Speech Recognition and
  Understanding (ASRU) Workshop}, 2019, pp. 1018--1025.

\bibitem{shon2018convolutional}
S.~Shon, A.~Ali, and J.~Glass, ``Convolutional neural network and language
  embeddings for end-to-end dialect recognition,'' in \emph{Proc. Odyssey 2018
  The Speaker and Language Recognition Workshop}, 2018, pp. 98--104.

\bibitem{KhuranaNAHBG17}
S.~Khurana, M.~Najafian, A.~M. Ali, T.~A. Hanai, Y.~Belinkov, and J.~R. Glass,
  ``{QMDIS:} {QCRI-MIT} advanced dialect identification system,'' in
  \emph{Interspeech}, F.~Lacerda, Ed., 2017, pp. 2591--2595.

\bibitem{AliM0R15}
A.~M. Ali, W.~Magdy, P.~Bell, and S.~Renals, ``Multi-reference {WER} for
  evaluating {ASR} for languages with no orthographic rules,'' in
  \emph{Automatic Speech Recognition and Understanding, {(ASRU)}}, 2015, pp.
  576--580.

\bibitem{miao2019new}
X.~Miao, I.~McLoughlin, and Y.~Yan, ``A new time-frequency attention mechanism
  for {TDNN} and {CNN-LSTM-TDNN}, with application to language
  identification,'' in \emph{Interspeech}, 2019, pp. 4080--4084.

\bibitem{miao2019lstm}
X.~Miao and I.~McLoughlin, ``{LSTM-TDNN} with convolutional front-end for
  dialect identification in the 2019 multi-genre broadcast challenge,''
  \emph{arXiv preprint arXiv:1912.09003}, 2019.

\bibitem{cai2019utterance}
W.~Cai, D.~Cai, S.~Huang, and M.~Li, ``Utterance-level end-to-end language
  identification using attention-based {CNN-BLSTM},'' in \emph{International
  Conference on Acoustics, Speech and Signal Processing (ICASSP)}, 2019, pp.
  5991--5995.

\bibitem{vaswani2017attention}
A.~Vaswani, N.~Shazeer, N.~Parmar, J.~Uszkoreit, L.~Jones, A.~N. Gomez,
  {\L}.~Kaiser, and I.~Polosukhin, ``Attention is all you need,'' in
  \emph{Advances in neural information processing systems}, 2017, pp.
  5998--6008.

\bibitem{li2019speechtransformer}
J.~Li, X.~Wang, Y.~Li \emph{et~al.}, ``The speechtransformer for large-scale
  mandarin chinese speech recognition,'' in \emph{International Conference on
  Acoustics, Speech and Signal Processing {(ICASSP)}}, 2019, pp. 7095--7099.

\bibitem{PoveyHGLK18}
D.~Povey, H.~Hadian, P.~Ghahremani, K.~Li, and S.~Khudanpur, ``A
  time-restricted self-attention layer for {ASR},'' in \emph{International
  Conference on Acoustics, Speech and Signal Processing, {ICASSP}}, 2018, pp.
  5874--5878.

\bibitem{transformer_transducer}
C.~Yeh, J.~Mahadeokar, K.~Kalgaonkar, Y.~Wang, D.~Le, M.~Jain, K.~Schubert,
  C.~Fuegen, and M.~L. Seltzer, ``Transformer-transducer: End-to-end speech
  recognition with self-attention,'' \emph{CoRR}, vol. abs/1910.12977, 2019.

\bibitem{PhamNN0W19}
N.~Pham, T.~Nguyen, J.~Niehues, M.~M{\"{u}}ller, and A.~Waibel, ``Very deep
  self-attention networks for end-to-end speech recognition,'' in
  \emph{Interspeech}, 2019, pp. 66--70.

\bibitem{SperberNNSW18}
M.~Sperber, J.~Niehues, G.~Neubig, S.~St{\"{u}}ker, and A.~Waibel,
  ``Self-attentional acoustic models,'' in \emph{Interspeech}, 2018, pp.
  3723--3727.

\bibitem{dong2018speech}
L.~Dong, S.~Xu, and B.~Xu, ``Speech-transformer: a no-recurrence
  sequence-to-sequence model for speech recognition,'' in \emph{International
  Conference on Acoustics, Speech and Signal Processing {(ICASSP)}}, 2018, pp.
  5884--5888.

\bibitem{SalazarKH19}
J.~Salazar, K.~Kirchhoff, and Z.~Huang, ``Self-attention networks for
  connectionist temporal classification in speech recognition,'' in
  \emph{International Conference on Acoustics, Speech and Signal Processing
  {(ICASSP)}}, 2019, pp. 7115--7119.

\bibitem{AlexBie}
A.~Bie, B.~Venkitesh, J.~Monteiro, M.~A. Haidar, and M.~Rezagholizadeh, ``Fully
  quantizing a simplified transformer for end-to-end speech recognition,''
  \emph{CoRR}, vol. abs/1911.03604, 2019.

\bibitem{TsunooKKW19}
E.~Tsunoo, Y.~Kashiwagi, T.~Kumakura, and S.~Watanabe, ``Transformer {ASR} with
  contextual block processing,'' in \emph{Automatic Speech Recognition and
  Understanding Workshop {(ASRU)}}, 2019, pp. 427--433.

\bibitem{sak2015fast}
H.~Sak, A.~Senior, K.~Rao, and F.~Beaufays, ``Fast and accurate recurrent
  neural network acoustic models for speech recognition,'' \emph{arXiv preprint
  arXiv:1507.06947}, 2015.

\bibitem{sak2015acoustic}
A.~W. Senior, H.~Sak, F.~de~Chaumont~Quitry, T.~N. Sainath, and K.~Rao,
  ``Acoustic modelling with {CD-CTC-sMBR LSTM RNNs},'' in \emph{Automatic
  Speech Recognition and Understanding (ASRU)}, 2015, pp. 604--609.

\bibitem{pundak2016lower}
G.~Pundak and T.~N. Sainath, ``Lower frame rate neural network acoustic
  models,'' in \emph{Interspeech}, 2016, pp. 22--26.

\bibitem{positional_encoding}
G.~Ke, D.~He, and T.~Liu, ``Rethinking positional encoding in language
  pre-training,'' \emph{CoRR}, vol. abs/2006.15595, 2020.

\bibitem{gehring2017convolutional}
J.~Gehring, M.~Auli, D.~Grangier, D.~Yarats, and Y.~N. Dauphin, ``Convolutional
  sequence to sequence learning,'' \emph{arXiv preprint arXiv:1705.03122},
  2017.

\bibitem{ShonASMG20}
S.~Shon, A.~M. Ali, Y.~Samih, H.~Mubarak, and J.~R. Glass, ``{ADI17:} {A}
  fine-grained arabic dialect identification dataset,'' in \emph{International
  Conference on Acoustics, Speech and Signal Processing {(ICASSP)}}, 2020, pp.
  8244--8248.

\bibitem{povey2011kaldi}
D.~Povey, A.~Ghoshal, G.~Boulianne, L.~Burget, O.~Glembek, N.~Goel,
  M.~Hannemann, P.~Motlicek, Y.~Qian, P.~Schwarz \emph{et~al.}, ``The kaldi
  speech recognition toolkit,'' in \emph{Automatic Speech Recognition and
  Understanding (ASRU)}.\hskip 1em plus 0.5em minus 0.4em\relax IEEE Signal
  Processing Society, 2011.

\end{thebibliography}



\end{document}